# Surface Tension of *Ab Initio* Liquid Water at the Water-Air Interface

*Yuki Nagata[1*] Tatsuhiko Ohto,[2] Mischa Bonn,[1] and Thomas D Kühne[3]*

1. Max Planck Institute for Polymer Research, Ackermannweg 10, 55128, Mainz, Germany
2. Graduate School of Engineering Science, Osaka University, 1-3 Machikaneyama, Toyonaka, Osaka, 560-8531, Japan
3. Dynamics of Condensed Matter, Department of Chemistry, University of Paderborn, Warburger Strasse 100, 33098 Paderborn, Germany

Corresponding authors: nagata@mpip-mainz.mpg.de

**ABSTRACT**:

We report calculations of the surface tension of the water-air interface using *ab initio* molecular dynamics (AIMD) simulations. We investigate the simulation cell size dependence of the surface tension of water from force field molecular dynamics (MD) simulations, which show that the calculated surface tension increases with increasing simulation cell size, thereby illustrating that a correction for finite size effects is required for the small system used in the AIMD simulation.

The AIMD simulations reveal that the double-$\xi$ basis set overestimates the experimentally measured surface tension due to the Pulay stress, while the triple and quadruple-$\xi$ basis sets give similar results. We further demonstrate that the van der Waals corrections critically affect the surface tension. AIMD simulations without the van der Waals correction substantially underestimate the surface tension, while van der Waals correction with the Grimme's D2 technique results in the value for the surface tension that is too high. The Grimme's D3 van der Waals correction provides a surface tension close to the experimental value. Whereas the specific choices for the van der Waals correction and basis sets critically affect the calculated surface tension, the surface tension is remarkably insensitive to the details of the exchange and correlation functionals, which highlights the impact of long-range interactions on the surface tension. These simulated values provide important benchmarks, both for improving van der Waals corrections, and AIMD simulations of aqueous interfaces.

I.         **INTRODUCTION:**

The surface tension of water governs the formation and shape of a water droplet, its contact angle when in contact with a surface, and its capillary action. The surface tension controls the wetting properties of water on material, as well as the ability of water striders to stay on the water surface. The surface tension of water at the water-air interface has been measured experimentally and was found to decrease substantially with increasing temperature: The surface tension is 72 mN/m at 300 K, reducing to 53 mN/m at 400 K.[1] It is also well-known that the surface tension of water decreases dramatically through the addition of surfactants on the water surface, while it increases by putting NaCl salt into the water.[2] These observation indicate that the surface tension is very sensitive with respect to water-water, water-surfactant, and water-ion interactions. On the other hand, the surface tensions of $H_2O$ and $D_2O$ at the water-air interface are rather similar,[3] implying that the surface tensions are rather insensitive to the variation of the hydrogen bond strength, known to be different for the two isotopes due to nuclear quantum effects. Since the surface tension of water is affected by the structure of the interfacial water molecules, nanoscale capillary waves, short- and long-range molecular interactions, and the surface activity of solutes, predicting the surface tension of water *in silico* is still very challenging.[4–9]

Computer simulations of the surface tension can provide molecular level insights into the relation between the surface tension of a macroscopic molecular liquid and microscopic molecular interactions.[10–19] In addition, calculating the surface tension of water at the water-vapor interface offers a critical check for empirical water models, since good agreement with experimentally measured surface tensions is a necessary condition for ensuring that the description of the aqueous interface is accurate. The surface tension of water at the water-vapor interface has been simulated, with rigid, non-polarizable water models such as SPC/E and TIP4P models,[11,20,21] as well as flexible, polarizable force field models such as AMOEBA[22] and POLI2VS models.[23] These models have been further employed to elucidate the nature of capillary waves of water,[21] the evaporation of water molecules,[24–26] the contact angle of a water droplet,[27–29] and vibrational spectra of water at the water-air interface.[23,30–33]

Recently, *ab initio* MD (AIMD) simulation techniques[34–36] have been used to elucidate the structure and dynamics of the hydrogen bond network,[37,38] and the acidity of interfacial water.[39–48] AIMD simulations at the aqueous interface have further been proven to be very useful for understanding surface-specific sum-frequency generation (SFG) vibrational spectra,[49,50] since the interpretation of the spectra does not depend on the employed water models. The imaginary parts of the SFG spectra by AIMD simulations at the water-air interface are marked by the absence of the positive band at 3200 $cm^{-1}$,[49] in agreement with recent experimental measurements.[51,52] Moreover, the SFG response of water at the water-lipid/surfactant interfaces revealed the preferential orientation of the water molecules in contact with the zwitterionic lipid/surfactant monolayer, again in good agreement with the experimental results.[53] Despite the increasing use of AIMD simulations to study aqueous interfaces, the surface tension of *ab initio* liquid water has, to the best of our knowledge, not yet been reported. The surface tension calculation constitutes a critical check for the AIMD description of aqueous interfaces.

Here, we present calculations of the surface tension of *ab initio* liquid water at the water-air interface, using the test-area method.[54] Since the AIMD simulations are computationally very demanding, and correspondingly can only be conducted using a relatively small simulation cell, we first investigate the size dependence of the surface tension at the water-air interface using classical force field models. We find that the tail correction of the LJ potential, as well as the finite size scaling method are well in computing the surface tension even for systems containing as few as 80 water molecules. Following this, multiple AIMD simulations of the water-air interface using different levels of theory have been performed. The reproducibility of the surface tension based on AIMD trajectories is discussed in detail.

The remaining of the paper is organized as follows. In Sec. II, we provide the simulation protocols of our force field based MD and AIMD simulations. Thereafter, we introduce the test-area method. In Sec. III, the simulated surface tensions from the force field MD and AIMD trajectories are reported and the impacts of the basis set, exchange and correlation (XC) functionals, and van der Waals correction on the surface tension are discussed. The conclusion is presented in Sec. IV.

## II. SIMULATION PROTOCOLS:

*Force Field MD Simulation*

To examine the simulation cell size dependence of the calculated surface tension at the water-air interface, we carried out classical force field MD simulations for various cell sizes. Since, in the AIMD simulations, the water molecules are inherently flexible, we employed the flexible SPC/Fw water model.[55] Periodic boundary conditions were employed for all $x$, $y$, and $z$-directions. The smooth particle mesh Ewald (SPME) method was used to compute the long-range electrostatic interactions. The cell size, Lennard-Jones (LJ) cutoff $r_c$, and the damping factor of the real part calculation in the Ewald summation $\alpha$,[56] were varied with the number of water molecules $N$, contained in the cell. In this study, we used $N$ = 80, 160, 320, 640, and 1280. The parameters of $N$, $r_c$, $\alpha$, and cell size ($L_x$, $L_y$, $L_z$) are summarized in Table 1, where the $xy$-plane is parallel to the water-air interface, and the $z$-axis forms the surface normal. Note that the real-space cutoff of the Ewald summation was set to be equal to the LJ cutoff. The equations of motion were integrated with the time step of 0.3 fs. All simulations have been performed using the CP2K software package.[57] We ran the MD simulations in the canonical *NVT* ensemble at a temperature of 300 K, which was controlled by the canonical sampling through the velocity rescaling (CSVR) approach.[58] The time constant of the thermostat was set to 1 ps. We obtained over 22.5 ns long MD trajectories for the systems with $L_x$ = 13.2, 16.63, and 20.95 Å, following 1 ns long MD runs to equilibrate the systems, while we obtained over 10.5 ns and 3.8 ns long MD trajectories for $L_x$ = 26.4 and 33.26 Å, respectively, again after 1 ns equilibration runs. From these 3.8 – 22.5 ns long MD trajectories, we sampled the snapshots every 1000 MD steps (every 0.3 ps), which were used for the surface tension calculation.

Table 1: Simulation parameters for surface tension calculation.

| Number of Molecules (*N*) | Cell Size ($V = L_x \times L_y \times L_z$ Å³) | LJ cutoff ($r_c$ Å) | Ewald ($\alpha$ Å$^{-1}$) |
|---|---|---|---|
| 80 | 13.20 × 13.20 × 35.00 | 6.00 | 0.583 |
| 160 | 16.63 × 16.63 × 44.10 | 7.56 | 0.463 |
| 320 | 20.95 × 20.95 × 55.56 | 9.52 | 0.368 |
| 640 | 26.40 × 26.40 × 70.00 | 12.00 | 0.292 |
| 1280 | 33.26 × 33.26 × 88.19 | 15.12 | 0.231 |

*Ab Initio MD Simulation*

To simulate the surface tensions by means of AIMD using various levels of theory, we have employed the Becke-Lee-Yang-Parr (BLYP)[59,60] XC functional/double-ξ valence polarized (DZVP) basis set with Grimme's D3 van der Waals correction[61] (BLYP/DZVP-D3), BLYP/triple-ξ valence plus two sets of polarization functions (TZV2P) basis set with the D3 correction (BLYP/TZVP-D3), BLYP/quadruple-ξ valence plus three sets of polarization functions (QZV3P) basis set with the D3 correction (BLYP/QZVP-D3), BLYP/TZVP with the Grimme's D2 van der Waals correction[62] (BLYP/TZVP-D2), BLYP/TZVP without van der Waals correction (BLYP/TZVP), Perdew-Burke-Ernzerhof[63] (PBE)/TZV2P with the D3 correction (PBE/TZVP-D3), and revised PBE[64] with the D3 correction (revPBE/TZVP-D3). The Goedecker-Teter-Hutter psuedopotential was used throughout to describe the core electrons,[65,66] together with an auxiliary plane wave cutoff of 320 Ry. To accelerate the MD simulation, we used $D_2O$ instead of $H_2O$ and a discretized time step of 0.5 fs. Note that the experimentally measured surface tensions of $H_2O$ and $D_2O$ are almost identical.[3] As before, the simulations were performed at $T$ = 300 K in the *NVT* ensemble with the CSVR thermostat. The number of the water molecules were set to $N$ = 80 and the cell size used in the AIMD simulations were $L_x$ = $L_y$ = 13.2 Å, and $L_z$ = 35.0 Å, respectively. A LJ cutoff of 6 Å was employed and the long-range corrections of the van der Waals interactions were turned off, in order to make the discussion consistent with the present force field simulations. After a 1 ns long classical force field MD run with the SPC/Fw water model, we conducted over 50 ps of AIMD simulation at the BLYP/DZVP-D3 level of theory in the *NVT* ensemble. In this step, we used a time constant for the CSVR thermostat of 50 fs. By using the coordinates and velocities obtained from the 50 ps BLYP/DZVP-D3 AIMD run, we performed 50 ps long AIMD runs at various levels of theory for equilibrating the systems, where we used a time constant for the CSVR thermostat of 1 ps. In this way, we obtained 440 ps (BLYP/DZVP-D3), 347 ps (BLYP/TZV2P-D3), 204 ps (BLYP/QZV3P-D3), 324 ps (BLYP/TZV2P), 289 ps (BLYP/TZV2P-D2), 245 ps (PBE/TZV2P-D3), and 264 ps (revPBE/TZV2P-D3) long AIMD trajectories in the *NVT* ensemble. The molecular geometries were sampled every 1000 AIMD steps (snapshots every 0.5 ps), and were used for the surface tension calculation.

*Test Area Method*

Surface tensions $\gamma$, were calculated via the test area method as[54]

$$\gamma_{sim} = -\frac{kT}{2\Delta S}\left(\ln\left\langle\exp\left(-\frac{\Delta U^+}{kT}\right)\right\rangle - \ln\left\langle\exp\left(-\frac{\Delta U^-}{kT}\right)\right\rangle\right), \quad (1)$$

where $\Delta U^{\pm}$ represents the conformational energy difference by scaling the cell size from $S = 2L_xL_y$ to $S \pm \Delta S$. In this simulation, we chose $\Delta S = 0.00005\ S$. This value is 10 times smaller than the value recommended for MD simulations with rigid-body force fields such as the SPC/E and TIP4P water models.[11] Since in the present simulation, the water molecules are made of flexible water monomers, the geometry of the water molecules can be scaled straightforwardly unlike the rigid-body models. Thus, the artifact induced by scaling the box size can be minimized, allowing us to use smaller a $\Delta S$ at variance to the rigid-body water models.[23]

To calculate the energy differences $\Delta U^{\pm}$, we used more accurate settings to calculate the electrostatic interactions. For the force field MD simulations, we used a finer grid with a spacing of (0.5 Å)$^3$ for the SPME calculation, while the other parameters were unchanged. For the AIMD simulations, we used a stricter self-consistent field convergence criterion for the $\Delta U^{\ddagger}$ calculation.

*Finite Size Corrections of Surface Tension*

We examined two ways for correcting finite size effects; the long-range tail correction for the LJ interactions[10,11,21] and the finite-size scaling technique of Binder and coworkers[63-66]. First, we outline the tail correction for the LJ interactions $\gamma_{tail}$, which can be estimated from

$$\gamma_{tail} = 12\pi\varepsilon\sigma^6(\rho_l - \rho_v)^2 \int_0^1 ds \int_{r_c}^{\infty} dr \coth(rs/d)\frac{3s^2 - s}{r^3}, \quad (2)$$

where $\rho_l$, $\rho_v$, and $d$ are the liquid density, vapor density, and the thickness parameter of the interfacial water region, while $\varepsilon$ and $\sigma$ are the parameters of the 12-6 LJ potential. In this study, we used the experimentally obtained density of water at 300 K ($\rho_l$ = 0.997 g/cm³), while $\rho_v$ is negligible at 300 K and thus was set to zero. The parameters $\varepsilon$ and $\sigma$ are given in the SPC/Fw water model.[55] The thickness of the interfacial water layer $d$, was calculated from

$$d = \sqrt{48\Delta^2/\pi^2}, \quad (3)$$

$$\Delta^2 = \Delta_0^2 + \frac{kT}{2\pi\gamma_0}\ln(L_x/b), \quad (4)$$

where $\Delta$ is the observed interfacial width, $\Delta_0$ the intrinsic interfacial width, $b$ the characteristic length scale associated with the short wavelength cutoff, and $\gamma_0$ the shear viscosity of liquid water.[21] In order to calculate $d$, we used $\Delta_0$ = 1.6 Å and $b$ = 4.8 nm,[67] as well as the experimentally measured surface tension of $\gamma_0$ = 72 mN/m.

Second, we outline the finite-size scaling technique. The size dependence of the simulated surface tension has been examined using Monte Carlo simulations, which has revealed that the simulated surface tension $\gamma_{sim}(L_x)$ has approximately $1/L_x$[68,69] or $\ln(L_z)/L_x^2$ dependence. [70,71] From the variation of the simulated surface tensions with varying cell sizes, we can extrapolate to the surface tension at infinite cell size. We calculate here the finite size effects from both the long-range LJ correction (Eq. (2)) and the extrapolation scheme.

## III. RESULTS:

*Size Effects of Surface Tension Simulated with SPC/Fw Water Model*

We calculated the surface tension using the SPC/Fw water model via the test-area method. The simulated surface tensions, $\gamma_{sim}$, are 49.8 ± 2.3, 55.0 ± 2.0, 56.5 ± 2.0, 59.2 ± 2.6, and 62.9 ± 3.8 mN/m for the systems characterized by $L_x$ = 13.20, 16.63, 20.95, 26.40, and 33.26 Å, respectively, where the error bars represent the respective 95 % confidential intervals (see Fig. 1). The surface tension clearly increases with *N*, which indicates that the surface tension depends strongly on the size of the simulation cell.

To estimate the finite size effects, we calculated the tail contributions of the LJ interactions from Eq. (2). The tail contributions were $\gamma_{tail}$ = 18.9, 12.2, 7.8, 5.0, and 3.1 mN/m, for $L_x$ = 13.20, 16.63, 20.95, 26.40, and 33.26 Å, respectively. The surface tensions corrected by Eq. (2) are also plotted in Fig. 1. As can be seen in this figure, the total surface tensions, $\gamma_{sim} + \gamma_{tail}$, are similar for different cell size within the error bar of ~5 mN/m, indicating that the LJ tail correction can correct reasonably well for the finite-size effects of water at the water-air interface. The simulated surface tensions using the SPC/Fw water model were 64.1 ± 2.6 mN/m for the $N$ = 640 system and 66.1 ± 3.8 mN/m for the $N$ = 1280 system, respectively.

Next, we estimated the finite size effects by extrapolating the surface tension to the infinite system using the finite size scaling technique of Binder and coworkers from a series of the simulated surface tensions with different cell sizes.[70,71] To that extent, we extrapolated the surface tension of the infinite system $\gamma_{sim}(L_x = \infty)$ from the simulated surface tension $\gamma_{sim}(L_x)$. Monte Carlo simulations,[68,69] as well as the analytical derivation[72] on the Ising model demonstrated that the surface tension has $1/L_x$ dependence. The fit with the $1/L_x$ dependence provide $\gamma_{sim}(L_x = \infty)$ = 70.1 ± 5.2 mN/m for the infinite cell size ($1/L_x \rightarrow 0$). However, a more recent study with larger simulation cells has revealed that the surface tension has a $\ln(L_z)/L_x^2$ dependence.[70,71] We also fit the data with the $\ln(L_z)/L_x^2$ dependence, which gives $\gamma_{sim}(L_x = \infty)$ = 64.9 ± 5.0 mN/m. Since the fit to the $\ln(L_z)/L_x^2$ dependence increases more slowly than that to $1/L_x$, in the region of small $1/L_x$, the fit with the $\ln(L_z)/L_x^2$ dependence provides a smaller $\gamma_{sim}(L_x = \infty)$. It should be noted that the surface tension calculated by $\gamma_{sim} + \gamma_{tail}$ agrees rather well with $\gamma_{sim}(L_x = \infty)$ = 64.9 mN/m as extrapolated using the $\ln(L_z)/L_x^2$ dependence, indicating that qualitatively, there is no large deviation between the extrapolation with the $\ln(L_z)/L_x^2$ dependence and the tail correction. By comparing this with the surface tension for the $N$ = 80 system, we found the finite size correction on the surface tension as $\gamma_{fse}$ (13.2Å $\rightarrow \infty$) = 15.1 mN/m, which again is rather similar to the LJ tail correction of $\gamma_{tail}$ = 18.9 mN/m for $L_x$ = 13.20 Å.

The simulated surface tension of ~65 mN/m from the SPC/Fw water model is slightly smaller than the experimentally measured surface tension of 72 mN/m. Note that this value is also smaller than the 70.8 ± 1.9 mN/m reported previously.[73] However, since the LJ and electrostatic interactions were both truncated beyond the cutoff length of 10 Å in Ref. [73], the different surface tensions between the current study and Ref. [73] can be attributed to the artificial truncation of the LJ and electrostatic forces.

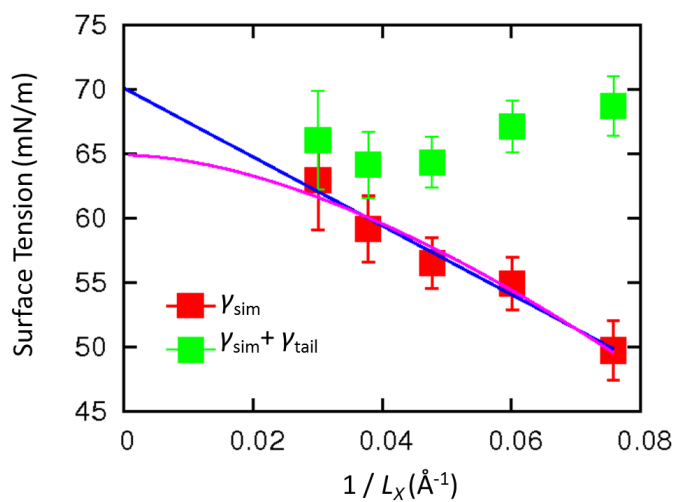

Figure 1. Simulated surface tensions using the SPC/Fw water model as a function of cell size. The blue and pink lines represent the fits with $1/L_x$ and $\ln(L_z)/L_x^2$ dependences, respectively.

*Surface Tensions for the Ab Initio Water-Air Interface*

From the AIMD simulation with the cell size of $L_x$ = 13.2 Å, we obtained surface tensions of $\gamma_{sim}$ = 152 ± 22 mN/m (BLYP/DZVP-D3), 92 ± 25 mN/m (BLYP/TZV2P-D3), 80 ± 34 mN/m (BLYP/QZV3P-D3), 20 ± 23 mN/m (BLYP/TZV2P), 126 ± 26 mN/m (BLYP/TZV2P-D2), 81 ± 29 mN/m (PBE/TZV2P-D3), and 83 ± 28 mN/m (revPBE/TZV2P-D3), respectively. These are also plotted in Fig. 2 including the aforementioned estimated contribution due to finite size effects of $\gamma_{fse}(13.2\text{Å} \rightarrow \infty)$ = 15.1 mN/m.

First, we compared the surface tensions by changing the Gaussian basis set (DZVP, TZV2P, and QZV3P) using the BLYP XC functional and Grimme's D3 van der Waals correction. The surface tension decreases dramatically from 152 to 80 mN/m with increasing accuracy of the employed basis set. Hence, the overestimated surface tension can be attributed to the Pulay stress, which results from the incompleteness of the employed basis set.[34,74] However, since in the CP2K/Quickstep code, a dual Gaussian and plane wave basis set is used,[75] there are two types of Pulay stresses. One is due to the incomplete Gaussian basis and another arises from the varying number of plane waves with respect to changes of the cell volume.[34] To quantify the dependence of the plane wave density cutoff on the surface tension, we recalculated the energy difference and then the surface tension with a rather large cutoff of 1200 Ry using the BLYP/DZVP+D3 AIMD snapshots. The computed surface tension with the cutoff of 1200 Ry was 152 mN/m, which is identical to the surface tension with the smaller cutoff of 320 Ry. This illustrates that the plane wave cutoff does not affect the surface tension, which entails that the large Pulay stress contribution is mainly due to the incomplete Gaussian basis set. This is presumably because *ΔS* is so small that the Pulay stress contribution from the auxiliary plane wave basis is negligible. Given the error bars, we conclude that the TZV2P and more accurate basis sets provide reasonable values for the surface tensions, while the DZVP basis set is clearly inadequate for the surface tension calculation.

Subsequently, we discuss the effects of the van der Waals correction on the simulated surface tension. Our data clearly demonstrate that the presence of the van der Waals correction improves the surface tension. Moreover, different types of van der Waals corrections provide rather different adjustments to the computed surface tension, as is evident from the comparison of the surface tensions with the D2 and D3 corrections; the simulation with the D2 correction overestimates the surface tension of water quite dramatically, while the D3 corrections provides rather reasonable surface tensions.

The D2 and D3 corrections differ in several points: For the pairwise potential, the D3 technique contains of the $r^{-6}$ term with a damping factor and an additional $r^{-8}$ term, while the D2 scheme comprises only the $r^{-6}$ term. In addition, the D3 correction includes a three-body interaction potential. To estimate the contribution of the three-body term in the D3 technique to the surface tension, we computed the energy difference $\Delta U^{\pm}$ at the BLYP/TZV2P-D3 level of theory by switching off the three-body term. Note that for the BLYP/TZV2P-D3 AIMD trajectories the three-body van der Waals contribution has been employed. The simulated surface tension without the three-body term was 95 mN/m, which is very similar to the surface tension of 92 mN/m with the three-body term. This indicates that the three-body term has limited impact on the surface tension, and that only the different pairwise potentials employed in the D2 and D3 techniques is the underlying cause for the difference in the computed surface tension. Our result implies that the accuracy of the van der Waals corrections critically affects the surface tension.

We then focused on the effects of the various XC functionals on the simulated surface tension. One can find that the BLYP, PBE, and revPBE XC functionals provide similar surface tensions within error bars, which is in contrast to the substantial impact of the different van der Waals corrections on the surface tension of the water. This is somehow surprising, especially when considering that the type of XC functionals has a sizable impact on the peak height of the oxygen-oxygen radial distribution, as does the type of van der Waals correction.[76,77] This implies that the long-range interactions critically affects the calculation of the surface tension.

Finally, we compared the simulated surface tensions with experimental data. The surface tension at the BLYP/TZV3P-D3 level including the contribution due to finite size effects of $\gamma_{fse}$ (13.2Å $\rightarrow \infty$) = 15 mN/m provides a total surface tension of $\gamma_{sim}$ ($L_x$ = 13.2Å) + $\gamma_{fse}$ (13.2Å $\rightarrow \infty$) = 107 mN/m, which is ~50 % larger than the experimental value of 72 mN/m. From the temperature-dependent surface tension data,[1] the difference of ~35 mN/m between the simulation and experiment corresponds to ~170 K difference. This indicates that even the D3 technique is not sufficient to adequately correct for all of the van der Waals interactions. Moreover, these results demonstrate that the simulation of the surface tension provides a critical test for improving van der Waals corrections,[78–82] van der Waals XC functionals,[83–86] and the long-range exchange correction method.[87]

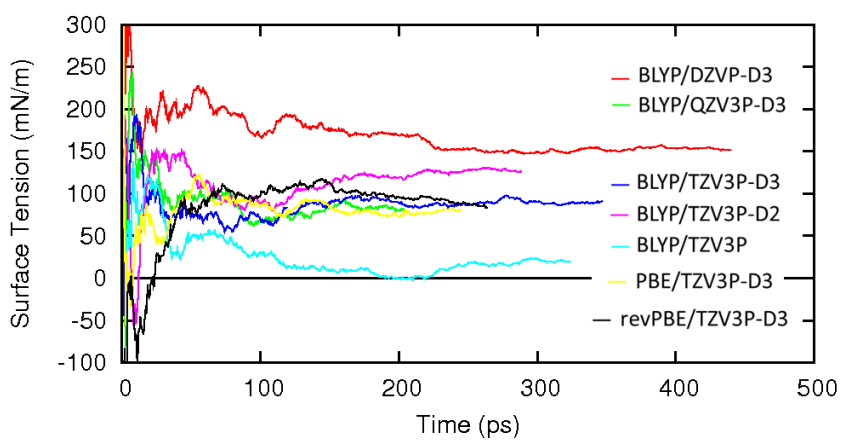

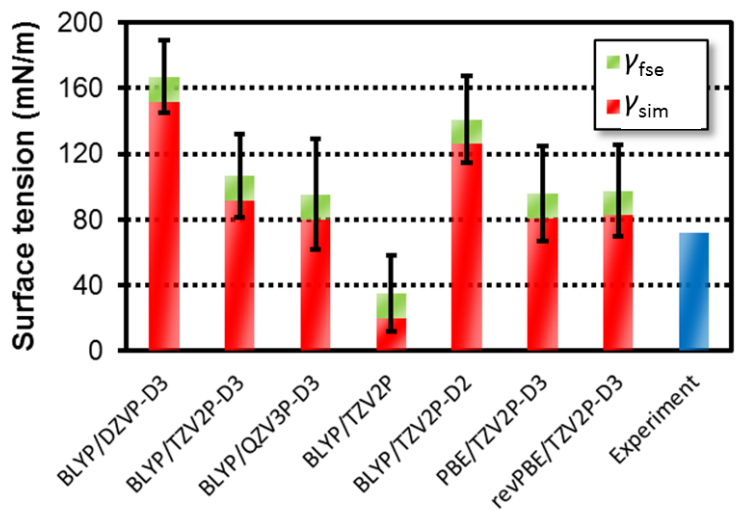

Figure 2. (Top) Time evolutions of the accumulated surface tensions as computed by AIMD simulations. (Bottom) Comparison of the computed surface tensions. The error bars represent 95 % confidential intervals.

IV.  **CONCLUSION:**

We have simulated the surface tension of the water-air interface by AIMD simulations. We find the simulated surface tensions depend critically on the employed basis sets, as well as the specific van der Waals correction, whereas they are much less sensitive to the used XC functionals. The usage of an imcomplete basis set leads to large Pulay stresses, leading to a dramatic overestimation of the surface tension (DVZP, 152 mN/m; TZV2P, 92 mN/m; QZV3P, 80 mN/m). Given the error bars, the TZV2P basis set can be recommended for computing the surface tension of water. Including van der Waals corrections substantially increases the surface tension from 20 mN/m to 126 mN/m for the D2 correction and to 92 mN/m for the D3 correction, respectively. This highlights the importance for delicate modeling of van der Waals interactions. High sensitivity of the van der Waals correction to the surface tension, in contrast to the insensitivity of the XC functionals, illustrates that the surface tension is governed by long-ranged interactions.

We also find that the surface tension is plagued by rather large finite size effects. The finite effects can be corrected via either the LJ tail correction or finite-size scaling technique with the $\ln(L_z)/L_x^2$ dependence. We confirm from the force field MD simulation that these provide similar finite size corrections. The computed surface tensions of water at the BLYP/TZV2P-D3 and BLYP/QZV3P give 107 and 95 mN/m, respectively, which are larger than the experimental value of 72 mN/m.

**ACKNOWLEDGEMENT**

We are grateful to Dr. Kurt Binder and Dr. Tristan Bereau for fruitful discussions.